\documentclass[12pt,draftcls,onecolumn]{IEEEtran}
\usepackage{makeidx,graphics,subfigure}
\usepackage{graphicx}
\usepackage{dcolumn}
\usepackage{bm}
\usepackage{multicol}
\usepackage{amsmath} 
\usepackage{amssymb}  

\begin{document}
%
\title{Manipulating quantum information on the controllable systems or subspaces}
\author{Ming Zhang\thanks{This work was funded by the National Natural
Science Foundation of China (Grant No.  60974037, 61074051 and
60821091), Ming Zhang and Jia-Hua Wei are with Department of Automatic Control,
College of Mechatronic Engineering and Automation, National
University of Defense Technology, Changsha, Hunan 410073, People's
Republic of China {\tt\small zhangming@nudt.edu.cn}}, Zairong Xi
IEEE Senior Member\thanks{ Zairong Xi is with Key Laboratory of
Systems and Control, AMSS, Chinese Academy of Sciences, Beijing, 100190,
China}, Jia-Hua Wei}
%
%
%
%
%
%
\markboth{IEEE Transition on Automatical Control,~Vol.~X,
No.~XX,~XXXX~20XX}{Shell \MakeLowercase{\textit{et al.}}: Bare Demo
of IEEEtran.cls for Journals}
%




 \maketitle

\begin{abstract}
In this paper, we explore how to constructively manipulate qubits by
rotating  Bloch spheres. It is revealed  that  three-rotation and
one-rotation Hamiltonian controls can be constructed to steer qubits
when two tunable Hamiltonian controls are available. It is
demonstrated in this research that local-wave function controls
 such as Bang-Bang, triangle-function and quadratic function controls can be
utilized to manipulate quantum states on the Bloch sphere. A new
kind of time-energy performance index is proposed
 to trade-off time and energy resource cost, in which
  control magnitudes are optimized in terms of
this kind of  performance.
It is further exemplified that this idea can be generalized to manipulate encoded qubits on the controllable subspace.
\end{abstract}


%
\IEEEpeerreviewmaketitle

\section{Introduction}
Dating from the birth of quantum theory, control of quantum systems
is an important issue \cite{Blaquiere1}. Quantum control
theory has been developed ever since 1980s\cite{Huang2,Ong
3,Clark4}. Recently, quantum information and quantum computation is the
focus of reseach\cite{quantum information}. A great progress has
been made in the domain of  quantum control\cite{quantum
control,dp}, in which the controllability of quantum systems is a
fundamental issue. The different notations of controllability have
been explored in \cite{R95,W06,S01,Z05,A03,W07,T01}. Specially, the
controllability of quantum open systems has been studied by some
researchers\cite{Schirmer 2003-19,Altafini 04-21,Lloyd 00-22,zm 06}.
It is quite well known that quantum open systems are not open-loop
controllable but there may exist decoherence-free subsystems or
subspaces\cite{42,43,44,45,46}. The works
on encoded universality\cite{47,48} further enhance the belief that
one can manipulate quantum information on the encoded subspace.
Optimal control theory has also been successfully applied to the
design of open-loop coherent control strategies  in physical
chemistry\cite{d12,d33,d34}. Recently, time-optimal control problems
for spin systems have been solved to achieve specified control
objectives in minimum time\cite{d35,d36,d37}. On the other hand, the
challenge of open-loop control is to design external fields or
potentials acting as model-relied controls. The
main strategies for open-loop control design
 seem to be based either on geometric ideas or more formally
Lie group decompositions, as in\cite{g1,g2,g3,g4,zszd}.

In this paper, we explore how to constructively manipulate qubits or
encoded qubits based on the geometric parametrization of qubits when two tunable Hamiltonian
controls are available. It is demonstrated that one can
not only design $3$-rotation Hamiltonian controls to manipulate
qubits, but can also
construct $1$-rotation Hamiltonian controls to steer qubits by
carefully choosing a rotation axis. It should be underlined that local
wave controls   can be constructed to manipulate qubits
corresponding to each rotation. Furthermore, we proposed  a new kind
of time-energy performance index
\begin{equation}
\label{J1}
J=\lambda\cdot{t_{f}}+\int_{0}^{t_{f}}E(u(t))dt=\int_{0}^{t_{f}}[\lambda+E(u(t))]dt
\end{equation}
where $E(u(t))$ is the energy cost of control  at time $t$, $t_{f}$
is free terminal time, and $\lambda$ is introduced as a ratio
parameter to trade-off the cost of time and energy resource. It has
also been discussed  in \cite{zszd} how to optimize  $3$-rotation
Bang-Bang controls to transfer quantum state in terms of this kind
of time-energy performance. In this paper, we comprehensively
discuss how to optimize  control magnitudes in terms of this kind of
time-energy performance for both $3$-rotation and $1$-rotation
controls, and present optimal Bang-Bang,  triangle-function and
quadratic function controls, respectively.

The rest of this paper are organized as follows. In Sect. II, we
present  prerequisite for further discussion. It is illustrated in
Sect. III how to manipulate qubit by $3$-rotation Bang-Bang,
triangle-function, and  quadratic function controls. The optimal
controls are further presented in the sense of time-energy
performance.  It is also revealed  in Sect. IV that one can utilize
three kinds of local-wave controls to manipulate  qubits just by  one-times rotation.
The paper concludes with Sect. V.

\section{Prerequisite}

Consider a controlled qubit governed by the equation
\begin{equation}
\label{2-1}
\frac{d}{dt}|\psi(t)\rangle=-\frac{{i}}{\hbar}H(t)|\psi(t)\rangle=-\frac{{i}}{\hbar}[u_{z}(t)\sigma_{z}+u_{y}(t)\sigma_{y}]|\psi(t)\rangle
\end{equation}
where $\sigma_{z}=I_{2}-2|1\rangle\langle1|$ and
$\sigma_{y}=i|1\rangle\langle0|-i|0\rangle\langle1|$. For
simplicity, we set $\hbar=1$.

Denote
$|u_{+}\rangle=\cos\frac{\theta_{u}}{2}|0\rangle+i\sin\frac{\theta_{u}}{2}|1\rangle;|u_{-}\rangle=\sin\frac{\theta_{u}}{2}|0\rangle-i\cos\frac{\theta_{u}}{2}|1\rangle$.
It is interesting to point out that if $H(t)=f(t)[\cos\theta_{u}\sigma_{z}+\sin\theta_{u}\sigma_{y}]$,
then one can express the Hamiltonian $H(t)$ as
$H(t)=f(t)\sigma^{u}_{z}$
where
$\sigma^{u}_{z}=|u_{+}\rangle\langle{u_{+}}|-|u_{-}\rangle\langle{u_{-}}|$.

\begin{figure}[ht]
\centering
\subfigure[$3$-rotation  trajectories]
 {\label{a}
\scalebox{0.35}{\includegraphics{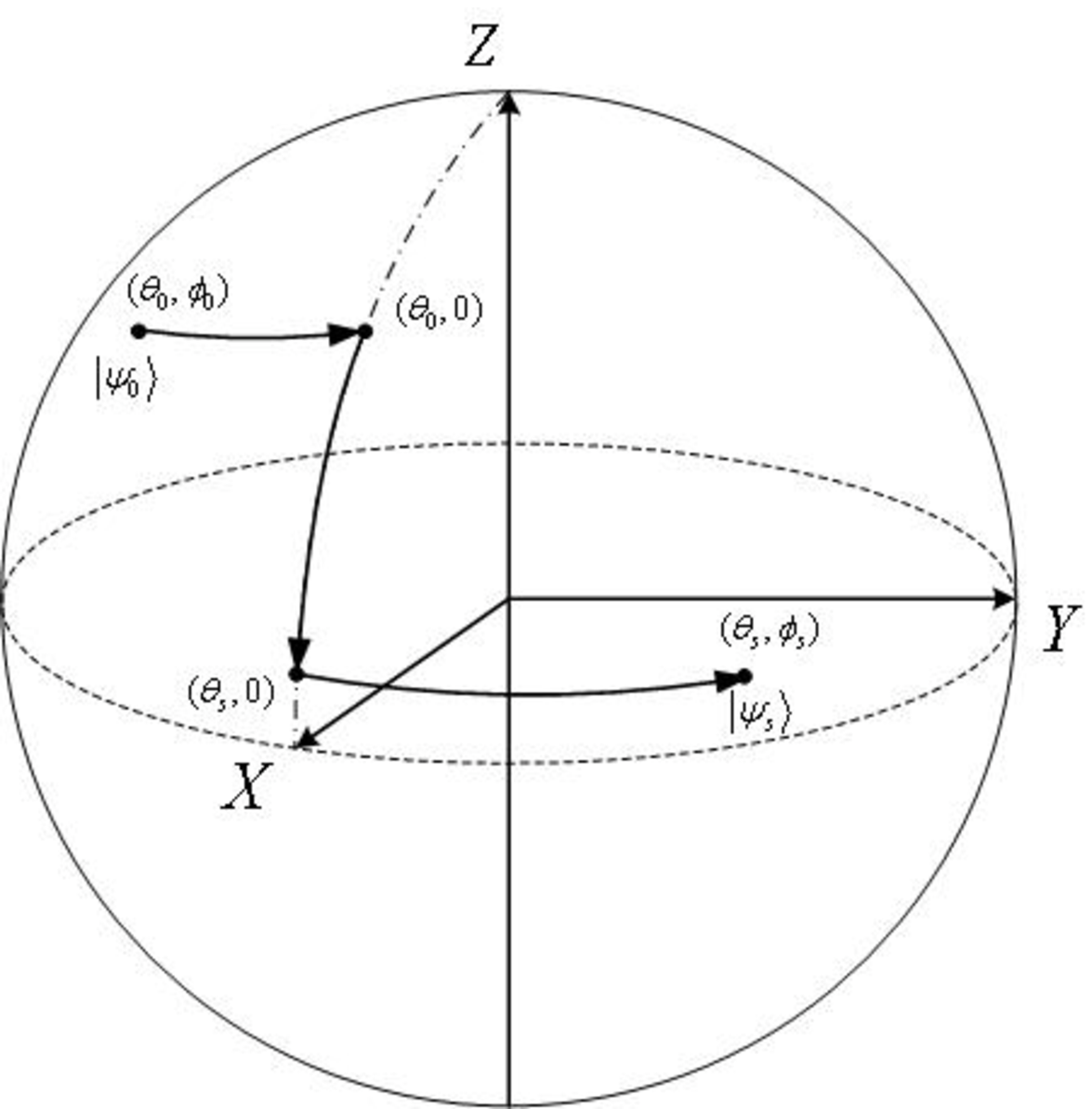}}
}%
\subfigure[$1$-rotation trajectory]
 {\label{b}
\scalebox{0.35}{\includegraphics{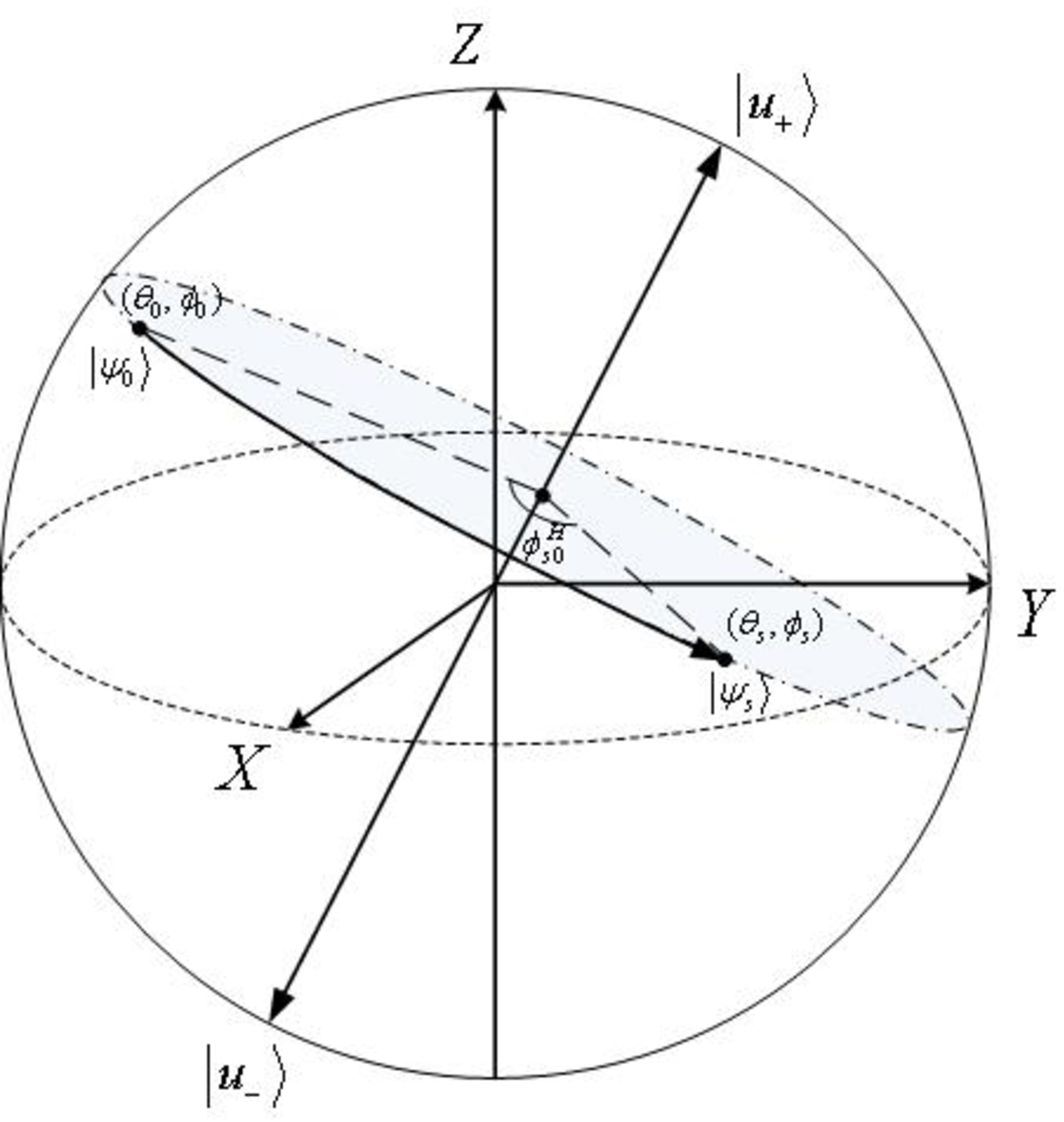}}
}%
\subfigure[$3$- and $1$-rotation trajectories]
 {\label{c}
\scalebox{0.35}{\includegraphics{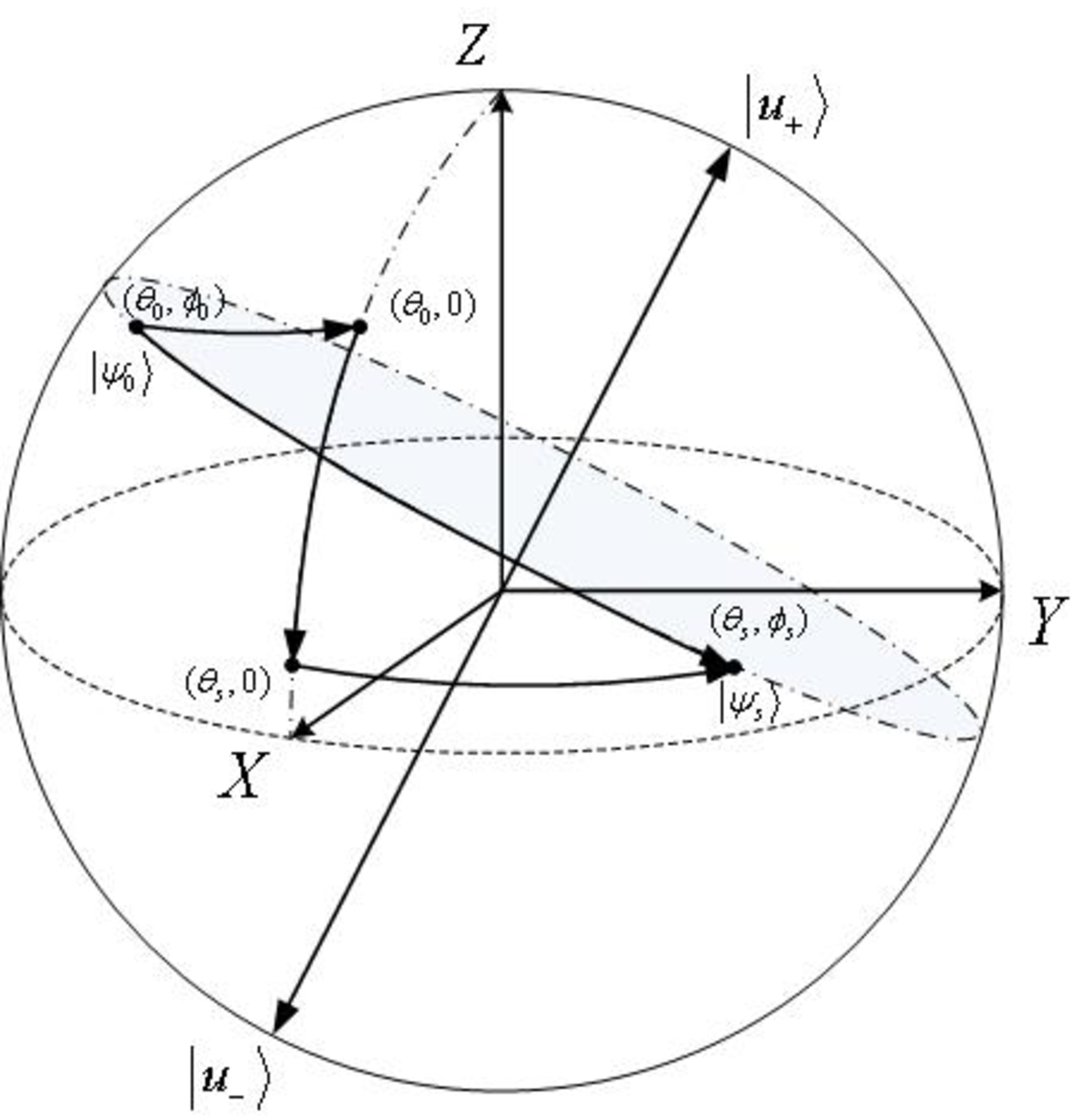}}
}%
\caption{\label{fig1} trajectories on the Bloch sphere}
\end{figure}

As shown in Fig.\ref{fig1}, one can not only choose $3$-rotation
control functions  to  steer the controlled qubit system from an
arbitrary initial state to another arbitrary target state, but can
also construct $1$-rotation control function $f(t)$ to achieve the
same goal.

In this paper, we will just
concentrate on three kind of local wave-functions: a piece-wise constant
function (Bang-Bang control), a triangle-function and a quadratic
function.

Denote the triangle function $u_{T}(t;t_{0},t_{1},L)$ and the quadratic
function $u_{Q}(t;t_{0},t_{1},L)$ respectively as follows:
\begin{equation}
\label{triangle} u_{T}(t;t_{0},t_{1},L)=\bigg\{\begin{array}{cc}
\frac{2L}{t_{1}-t_{0}}\cdot{(t-t_{0})}&t\in[t_{0},\frac{t_{0}+t_{1}}{2})\\
-\frac{2L}{t_{1}-t_{0}}\cdot{(t-t_{1})}&t\in[\frac{t_{0}+t_{1}}{2},t_{1})\\
0&otherwise
\end{array}
\end{equation}
and
\begin{equation}
\label{Sq} u_{Q}(t;t_{0},t_{1},L)=\bigg\{\begin{array}{cc}
\frac{4L\cdot{(t-t_{0})(t_{1}-t)}}{(t_{1}-t_{0})^{2}}&t\in[t_{0},t_{1}]\\
0&otherwise
\end{array}
\end{equation}
where both $u_{T}(t;t_{0},t_{1},L)$ and $u_{Q}(t;t_{0},t_{1},L)$ are
nonzero only when $t\in(t_{0},t_{1})$, and take  the maximum
magnitude $L$ at time $\frac{t_{0}+t_{1}}{2}$. It should be
underlined that the pulse area of the control pulses  is the key
control variable for geometric control and the pulse area inequality
for Bang-Bang, triangle-function and quadratic function controls is
given as
\begin{equation}\label{Area-i}
\int^{t_{1}}_{t_{0}}u_{T}(t;t_{0},t_{1},L)dt<\int^{t_{1}}_{t_{0}}u_{Q}(t;t_{0},t_{1},L)dt<\int^{t_{1}}_{t_{0}}Ldt=L(t_{1}-t_{0})
\end{equation}
Furthermore, it is worth pointing out that
$\int^{t_{1}}_{t_{0}}L^{2}dt=L^{2}(t_{1}-t_{0})$ and
\begin{equation}\label{tri-e}
E(u_{T}(t;t_{0},t_{1},L))=\int^{t_{1}}_{t_{0}}|u_{T}(t;t_{0},t_{1},L)|^{2}dt
=\frac{1}{3}L^{2}(t_{1}-t_{0})
\end{equation}
and
\begin{equation}\label{Q-e}
E(u_{Q}(t;t_{0},t_{1},L))=\int^{t_{1}}_{t_{0}}|u_{Q}(t;t_{0},t_{1},L)|^{2}dt
=\frac{8}{15}L^{2}(t_{1}-t_{0})
\end{equation}

\emph{Remark:} We would like to further emphasize that that one can
construct both $3$-rotation and $1$-rotation local wave-function controls
to manipulate qubits if $H(t)=u_{x}(t)\sigma_{x}+u_{y}(t)\sigma_{y}$
or $H(t)=u_{z}(t)\sigma_{z}+u_{x}(t)\sigma_{x}$. In other words,
$3$-rotation and $1$-rotation  controls can be
constructed as long as two tunable Hamiltonian controls are
available.

\section{Manipulate  qubit by three rotations}

Consider a controlled qubit governed by the equation
\begin{equation}
\label{2-14}
\frac{d}{dt}|\psi(t)\rangle=-{i}(u_{z}(t)\sigma_{z}+u_{y}(t)\sigma_{y})|\psi(t)\rangle
\end{equation}
with an initial state
$|\psi_{0}\rangle=\cos\frac{\theta_{0}}{2}|0\rangle+e^{i\phi_{0}}\sin\frac{\theta_{0}}{2}|1\rangle$
and a target state
$|\psi_{s}\rangle=\cos\frac{\theta_{s}}{2}|0\rangle+e^{i\phi_{s}}\sin\frac{\theta_{s}}{2}|1\rangle$.
For the sake of the following analysis in this section, we denote $\phi_{0m}=\min(\phi_{0},2\pi-\phi_{0})$, $\phi_{sm}=\min(\phi_{s},2\pi-\phi_{s})$,  $\theta_{0s}=|\theta_{s}-\theta_{0}|$ and $\Sigma_{\phi\theta}=\phi_{0m}+\theta_{0s}+\phi_{sm}$.

In this section,  our control goal is to find $t_{f}$ and  some form
of controls $\{u_{z}(t),u_{y}(t):0\leq{t}\leq{t_{f}}\}$ so that
\begin{equation}
\label{TE} |\psi(t_{f})\rangle=|\psi_{s}\rangle
\end{equation}
by three rotations about $z-$axis, $y-$axis and  $z-$axis, respectively. Furthermore, we hope to optimize control magnitude  in terms of
 the performance (\ref{J1})
where $E(u(t))=|u_{z}(t)|^{2}+|u_{y}(t)|^{2}$ and $\lambda>0$.

\subsection{$3$-rotation Bang-Bang controls}
In this subsection, we will discuss how to manipulate quantum system
(\ref{2-14}) by three-rotation Bang-Bang control. According to the
properties of Pauli matrices\cite{quantum information}, we choose
the piecewise constant controls
$\{u_{z}(t),u_{y}(t):0\leq{t}\leq{t_{f}}\}$ as follows:
\begin{equation}
\label{2-17} u_{z}(t)=\bigg\{\begin{array}{cc}
sign(\phi_{0}-\pi)M_{z1}&t\in[0,t_{1})\\
0&t\in[t_{1},t_{2})\\
sign(\pi-\phi_{s})M_{z2}&t\in[t_{2},t_{f}]
\end{array}
\end{equation}
and
\begin{equation}
\label{2-18} u_{y}(t)=\bigg\{\begin{array}{cc}
0&t\in[0,t_{1})\\
sign(\theta_{s}-\theta_{0})M_{y}&t\in[t_{1},t_{2})\\
0&t\in[t_{2},t_{f}]
\end{array}
\end{equation}
where $t_{1}=\frac{\phi_{0m}}{2M_{z1}}$,
$t_{2}=\frac{\theta_{0s}}{2M_{y}}+t_{1}$ and
$t_{f}=\frac{\phi_{sm}}{2M_{z2}}+t_{2}$.

After some calculations, we have
$|\psi(t_{1})\rangle=\cos\frac{\theta_{0}}{2}|0\rangle+\sin\frac{\theta_{0}}{2}|1\rangle$,
$|\psi(t_{2})\rangle=\cos\frac{\theta_{s}}{2}|0\rangle+\sin\frac{\theta_{s}}{2}|1\rangle$,
and
$|\psi(t_{f})\rangle=\cos\frac{\theta_{s}}{2}|0\rangle+e^{i\phi_{s}}\sin\frac{\theta_{s}}{2}|1\rangle$.

Next, our task is to choose $M_{z1}$, $M_{z2}$ and $M_{y}$ to
minimize
 the performance (\ref{J1}).
It can be demonstrated that
\begin{equation}
\label{2-19}\begin{array}{c}
J=\lambda(\frac{\phi_{0m}}{2M_{z1}}+\frac{\theta_{0s}}{2M_{y}}+\frac{\phi_{sm}}{2M_{z2}})
+(\frac{M_{z1}\phi_{0m}}{2}+\frac{M_{y}\theta_{0s}}{2}+\frac{M_{z2}\phi_{sm}}{2})
\geq\sqrt{\lambda}\Sigma_{\phi\theta}
\end{array}
\end{equation}
where the equality holds only if
$M_{z1}=M_{z2}=M_{y}=\sqrt{\lambda}$.

If only bounded Bang-Bang controls with bound $L_{B}$ are
permitted, then the optimal controls are given as:
\begin{equation}
\label{BO1} u^{*}_{z}(t)=\bigg\{\begin{array}{cc}
sign(\phi_{0}-\pi)L_{B}^{*}&t\in[0,t^{*}_{1})\\
0&t\in[t^{*}_{1},t^{*}_{2})\\
sign(\pi-\phi_{s})L_{B}^{*}&t\in[t^{*}_{2},t^{*}_{f}]
\end{array}
\end{equation}
and
\begin{equation}
\label{BO2} u^{*}_{y}(t)=\bigg\{\begin{array}{cc}
0&t\in[0,t^{*}_{1})\\
sign(\theta_{s}-\theta_{0})L_{B}^{*}&t\in[t^{*}_{1},t^{*}_{2})\\
0&t\in[t^{*}_{2},t^{*}_{f}]
\end{array}
\end{equation}
where $t^{*}_{1}=\frac{\phi_{0m}}{2L_{B}^{*}}$,
$t^{*}_{2}=\frac{\theta_{0s}}{2L_{B}^{*}}+t^{*}_{1}$,
$t^{*}_{fB}=\frac{\phi_{sm}}{2L_{B}^{*}}+t^{*}_{2}=\frac{\Sigma_{\phi\theta}}{2L_{B}^{*}}$
and $L_{B}^{*}=\min(\sqrt{\lambda},L_{B})$. Furthermore, the
corresponding optimal performance is
$J_{B}^{*}=(\frac{\lambda}{2L_{B}^{*}}+\frac{L_{B}^{*}}{2})\Sigma_{\phi\theta}$.
It is interesting to underline that
$J_{B}^{*}=\lambda\cdot{t^{*}_{fB}}+E^{*}_{B}$ where
$E^{*}_{B}=\frac{1}{2}L_{B}^{*}\Sigma_{\phi\theta}$, and
${t^{*}_{fB}}\cdot{E^{*}_{B}}=\frac{1}{4}\Sigma_{\phi\theta}^{2}$
only depends on the location of both initial and target states on
the Bloch sphere.

If unbounded Bang-Bang controls are permitted, we have $L_{B}^{*}=\sqrt{\lambda}$, $t^{*}_{fB}=\frac{\Sigma_{\phi\theta}}{2\sqrt{\lambda}}$,  $E^{*}_{B}=\frac{\sqrt{\lambda}}{2}\Sigma_{\phi\theta}$ and  $J_{B}^{*}=\sqrt{\lambda}\Sigma_{\phi\theta}$, therefore ${t^{*}_{fB}}\cdot{E^{*}_{B}}=\frac{1}{4}\Sigma_{\phi\theta}^{2}$.

\subsection{$3$-rotation triangle-function controls}
In this subsection, we will first demonstrate that  the target state
$|\psi(t_{f})\rangle=|\psi_{s}\rangle$
can be achieved from the initial state $|\psi_{o}\rangle$ by the following three-rotation triangle-function controls:
\begin{equation}
\label{tri2-17}
u_{z}(t)=sign(\phi_{0}-\pi)u_{T}(t;0,t_{1},M_{z1})+sign(\pi-\phi_{s})u_{T}(t;t_{2},t_{f},M_{z2})
\end{equation}
and
\begin{equation}
\label{tri2-18}
u_{y}(t)=sign(\theta_{s}-\theta_{0})u_{T}(t;t_{1},t_{2},M_{y})
\end{equation}
where $t_{1}=\frac{\phi_{0m}}{M_{z1}}$,
$t_{2}=\frac{\theta_{0s}}{M_{y}}+t_{1}$ and
$t_{f}=\frac{\phi_{sm}}{M_{z2}}+t_{2}$.
It can be proved  that
$|\psi(t_{1})\rangle=\cos\frac{\theta_{0}}{2}|0\rangle+\sin\frac{\theta_{0}}{2}|1\rangle$,
$|\psi(t_{2})\rangle=\cos\frac{\theta_{s}}{2}|0\rangle+\sin\frac{\theta_{s}}{2}|1\rangle$,
and
$|\psi(t_{f})\rangle=\cos\frac{\theta_{s}}{2}|0\rangle+e^{i\phi_{s}}\sin\frac{\theta_{s}}{2}|1\rangle$.

Subsequently, our task is to select magnitude $M_{z1}$, $M_{z2}$ and $M_{y}$
to minimize
 the performance (\ref{J1}).
It can be demonstrated that
\begin{equation}
\label{Tri2-19}\begin{array}{c}
J=\lambda(\frac{\phi_{0m}}{M_{z1}}+\frac{\theta_{0s}}{M_{y}}+\frac{\phi_{sm}}{M_{z2}})+(\frac{M_{z1}\phi_{0m}}{3}+\frac{M_{y}\theta_{0s}}{3}+\frac{M_{z2}\phi_{sm}}{3})
\geq\frac{2\sqrt{\lambda}}{\sqrt{3}}\Sigma_{\phi\theta}
\end{array}
\end{equation}
where the equality holds only if
$M_{z1}=M_{z2}=M_{y}=\sqrt{3\lambda}$.
If only bounded triangle-function controls with bound $L_{B}$
are permitted, then the optimal $3$-rotation  triangle-function controls are given as:
\begin{equation}
\label{triuo1}
u^{*}_{z}(t)=sign(\phi_{0}-\pi)u_{T}(t;0,t^{*}_{1},L_{T}^{*})+sign(\pi-\phi_{s})u_{T}(t;t^{*}_{2},t^{*}_{fT},L_{T}^{*})
\end{equation}
and
\begin{equation}
\label{triuo2}
u^{*}_{y}(t)=sign(\theta_{s}-\theta_{0})u_{T}(t;t^{*}_{1},t^{*}_{2},L_{T}^{*})
\end{equation}
where $t^{*}_{1}=\frac{\phi_{0m}}{L_{T}^{*}}$,
$t^{*}_{2}=\frac{\theta_{0s}}{L_{T}^{*}}+t^{*}_{1}$,
$t^{*}_{fT}=\frac{\Sigma_{\phi\theta}}{L_{T}^{*}}$ and
$L_{T}^{*}=\min(\sqrt{3\lambda},L_{B})$.  Furthermore, the optimal
performance corresponding to bounded  triangle-function  control is
$J_{T}^{*}=(\frac{\lambda}{L_{T}^{*}}+\frac{L_{T}^{*}}{3})\Sigma_{\phi\theta}$.
It is interesting to underline that
$J_{T}^{*}=\lambda\cdot{t^{*}_{fT}}+E^{*}_{T}$ with
$E^{*}_{T}=\frac{1}{3}L_{T}^{*}\Sigma_{\phi\theta}$, and
${t^{*}_{fT}}\cdot{E^{*}_{T}}=\frac{1}{3}\Sigma_{\phi\theta}^{2}$
only depends on the location of both initial and target states on
the Bloch sphere.

If unbounded triangle-function controls are permitted, then we have
 $t^{*}_{fT}=\frac{\Sigma_{\phi\theta}}{\sqrt{3\lambda}}$, $L_{T}^{*}=\sqrt{3\lambda}$, $E^{*}_{T}=\frac{\sqrt{3\lambda}}{3}\Sigma_{\phi\theta}$ and
$J_{T}^{*}=\frac{2\sqrt{\lambda}}{\sqrt{3}}\Sigma_{\phi\theta}$, thus ${t^{*}_{fT}}\cdot{E^{*}_{T}}=\frac{1}{3}\Sigma_{\phi\theta}^{2}$.

\subsection{$3$-rotation quadratic function controls}
In this subsection, it is  demonstrated that  the target state
$|\psi(t_{f})\rangle=|\psi_{s}\rangle$
can be achieved from the initial state $|\psi_{o}\rangle$ by the following
 quadratic controls:
\begin{equation}
\label{Sq2-17}
u_{z}(t)=sign(\phi_{0}-\pi)u_{Q}(t;0,t_{1},M_{z1})+sign(\pi-\phi_{s})u_{Q}(t;t_{2},t_{f},M_{z2})
\end{equation}
and
\begin{equation}
\label{Sq2-18}
u_{y}(t)=sign(\theta_{s}-\theta_{0})u_{Q}(t;t_{1},t_{2},M_{y})
\end{equation}
where $t_{1}=\frac{3\phi_{0m}}{4M_{z1}}$,
$t_{2}=\frac{3\theta_{0s}}{4M_{y}}+t_{1}$ and
$t_{f}=\frac{3\phi_{sm}}{4M_{z2}}+t_{2}$.
It can be confirmed  that
$|\psi(t_{1})\rangle=\cos\frac{\theta_{0}}{2}|0\rangle+\sin\frac{\theta_{0}}{2}|1\rangle$,
$|\psi(t_{2})\rangle=\cos\frac{\theta_{s}}{2}|0\rangle+\sin\frac{\theta_{s}}{2}|1\rangle$,
and
$|\psi(t_{f})\rangle=\cos\frac{\theta_{s}}{2}|0\rangle+e^{i\phi_{s}}\sin\frac{\theta_{s}}{2}|1\rangle$.

Next, our task is to choose magnitude $M_{z1}$, $M_{z2}$ and $M_{y}$
to minimize
 the performance (\ref{J1}).

After some calculations, we have
\begin{equation}
\label{Sqi2-19}\begin{array}{c}
J=\lambda(\frac{3\phi_{0m}}{4M_{z1}}+\frac{3\theta_{0s}}{4M_{y}}+\frac{3\phi_{sm}}{4M_{z2}})
+(\frac{2M_{z1}\phi_{0m}}{5}+\frac{2M_{y}\theta_{0s}}{5}+\frac{2M_{z2}\phi_{sm}}{5})
\geq\frac{{\sqrt{30\lambda}}}{5}\Sigma_{\phi\theta}
\end{array}
\end{equation}
where the equality holds only if
$M_{z1}=M_{z2}=M_{y}=\frac{\sqrt{30\lambda}}{4}$.

If only bounded quadratic function controls with bound $L_{B}$
are permitted, the optimal $3$-rotation bounded quadratic-function controls are given as:
\begin{equation}
\label{Squo1}
u^{*}_{z}(t)=sign(\phi_{0}-\pi)u_{Q}(t;0,t^{*}_{1},L_{Q}^{*})+sign(\pi-\phi_{s})u_{Q}(t;t^{*}_{2},t^{*}_{f},L_{Q}^{*})
\end{equation}
and
\begin{equation}
\label{Squo2}
u^{*}_{y}(t)=sign(\theta_{s}-\theta_{0})u_{Q}(t;t^{*}_{1},t^{*}_{2},L_{Q}^{*})
\end{equation}
where $t^{*}_{1}=\frac{3\phi_{0m}}{4L_{Q}^{*}}$,
$t^{*}_{2}=\frac{3\theta_{0s}}{4L_{Q}^{*}}+t_{1}$,
$t^{*}_{fQ}=\frac{3\Sigma_{\phi\theta}}{4L_{Q}^{*}}$ and
$L_{Q}^{*}=\min(\frac{\sqrt{30\lambda}}{4},L_{B})$. Moreover, the
optimal performance corresponding to bounded  control is
$J_{Q}^{*}=(\frac{3\lambda}{4L_{Q}^{*}}+\frac{2L_{Q}^{*}}{5})\Sigma_{\phi\theta}$.
It is interesting to underline that
$J_{Q}^{*}=\lambda\cdot{t^{*}_{fQ}}+E^{*}_{Q}$ where
$E^{*}_{Q}=\frac{2}{5}L_{Q}^{*}\Sigma_{\phi\theta}$, and
${t^{*}_{fQ}}\cdot{E^{*}_{Q}}=\frac{3}{10}\Sigma_{\phi\theta}^{2}$
only depends on the location of both initial and target states on
the Bloch sphere.

If unbounded quadratic function controls are permitted, we have
$t^{*}_{fQ}=\frac{3\Sigma_{\phi\theta}}{\sqrt{30\lambda}}$, $L_{Q}^{*}=\frac{\sqrt{30\lambda}}{4}$, $E^{*}_{Q}=\frac{\sqrt{30\lambda}}{10}\Sigma_{\phi\theta}$ and
$J_{Q}^{*}=\frac{\sqrt{30\lambda}}{5}\Sigma_{\phi\theta}$, therefore ${t^{*}_{fQ}}\cdot{E^{*}_{Q}}=\frac{3}{10}\Sigma_{\phi\theta}^{2}$.

\emph{Remark:}
1. When unbounded controls are permitted,  it has been  demonstrated  in this section that
$J^{*}_{B}<J^{*}_{Q}<J^{*}_{T}$, $E^{*}_{B}<E^{*}_{Q}<E^{*}_{T}$ and
 $t^{*}_{fB}<t^{*}_{fQ}<t^{*}_{fT}$, therefore we have $t^{*}_{fB}{\cdot}E^{*}_{B}<t^{*}_{fQ}{\cdot}E^{*}_{Q}<t^{*}_{fT}{\cdot}E^{*}_{T}$.

2. Even when only bounded controls are permitted, the
above inequalities are valid for all $\lambda$ and $L_{B}$
except that $E^{*}_{B}<E^{*}_{Q}<E^{*}_{T}$
does not hold for some
$\lambda$ and $L_{B}$.

\section{Manipulate qubits just by one rotation}

Reconsider the controlled qubit (\ref{2-14}) with both the same
initial and  target states given in the Section III. In this
section, our control goal is to find $t_{f}$ and  some form of
controls $\{u_{z}(t),u_{y}(t):0\leq{t}\leq{t_{f}}\}$ so that
$|\psi(t_{f})\rangle=|\psi_{s}\rangle$ is attained just by one
rotation. Furthermore, we hope to choose
$\{u_{z}(t),u_{y}(t):0\leq{t}\leq{t_{f}}\}$ to minimize
 the performance  (\ref{J1}).

Choose
$H(t)=f(t)(\cos\theta_{u}\sigma_{z}+\sin\theta_{u}\sigma_{y})$ with
$\theta_{u}\in[0,\pi]$ so that the following equation holds
\begin{equation}
\label{UEone}
\sin\theta_{u}\sin\theta_{0}\sin\phi_{0}+\cos\theta_{u}\cos\theta_{0}=\sin\theta_{u}\sin\theta_{s}\sin\phi_{s}+\cos\theta_{u}\cos\theta_{s}
\end{equation}
Since $|0\rangle=\cos\frac{\theta_{u}}{2}|u_{+}\rangle+i\sin\frac{\theta_{u}}{2}|u_{-}\rangle;|1\rangle=-i\sin\frac{\theta_{u}}{2}|u_{+}\rangle+i\cos\frac{\theta_{u}}{2}|u_{-}\rangle$,  the initial  and
target states  can be expressed
in terms of the new basis
$|u_{+}\rangle$ and $|u_{-}\rangle$ as follows
\begin{equation}
|\psi_{0}\rangle=\cos\frac{\theta^{H}_{s0}}{2}|u_{+}\rangle+e^{i\phi^{H}_{0}}\sin\frac{\theta^{H}_{s0}}{2}|u_{-}\rangle
\end{equation}
and
\begin{equation}
|\psi_{s}\rangle=\cos\frac{\theta^{H}_{s0}}{2}|u_{+}\rangle+e^{i\phi^{H}_{s}}\sin\frac{\theta^{H}_{s0}}{2}|u_{-}\rangle
\end{equation}
where
\begin{equation}
\label{UEone1}
\cos\frac{\theta^{H}_{s0}}{2}=\sqrt{\frac{1}{2}+\frac{1}{2}[\sin\theta_{u}\sin\theta_{0}\sin\phi_{0}+\cos\theta_{u}\cos\theta_{0}]}
\end{equation}
and
\begin{equation}
\label{UEone2}
\begin{array}{c}
\phi^{H}_{0}=-\angle(\cos\frac{\theta_{0}}{2}\cos\frac{\theta_{u}}{2}-ie^{i\phi_{0}}\sin\frac{\theta_{0}}{2}\sin\frac{\theta_{u}}{2})\\
+\angle(i\cos\frac{\theta_{0}}{2}\sin\frac{\theta_{u}}{2}+ie^{i\phi_{0}}\sin\frac{\theta_{0}}{2}\cos\frac{\theta_{u}}{2})\pm{2n_{0}}\pi
\end{array}
\end{equation}
and
\begin{equation}
\label{UEone3}
\begin{array}{c}
\phi^{H}_{s}=-\angle(\cos\frac{\theta_{s}}{2}\cos\frac{\theta_{u}}{2}-ie^{i\phi_{s}}\sin\frac{\theta_{s}}{2}\sin\frac{\theta_{u}}{2})\\
+\angle(i\cos\frac{\theta_{s}}{2}\sin\frac{\theta_{u}}{2}+ie^{i\phi_{s}}\sin\frac{\theta_{s}}{2}\cos\frac{\theta_{u}}{2})\pm{2n_{s}}\pi
\end{array}
\end{equation}
It is easy to prove that one can choose the suitable integers
$n_{0}$ and $n_{s}$ so that $\phi^{H}_{0},\phi^{H}_{s}\in[0,2\pi)$.

\emph{Remark:} 1. We would like to point out that  the initial  and
target states have the same angle $\theta^{H}_{s0}$ about the
control Hamiltonian axis as shown in Fig.\ref{fig1}.

2.  For the sake of the analysis,  we introduce
$\phi^{H}_{s0}=\min(|\phi^{H}_{s}-\phi^{H}_{0}|,2\pi-|\phi^{H}_{s}-\phi^{H}_{0}|)$ with $\phi^{H}_{s0}\in[0,\pi)$. It should be underlined that $\phi^{H}_{s0}$ depends not only on the location of both initial and target states on the Bloch sphere, but also on the Hamiltonian $H(t)$, i.e.,  the $y-z$ plane.

\subsection{$1$-rotation Bang-Bang controls}
In this subsection, we will discuss how to manipulate the quantum
system (\ref{2-14}) by Bang-Bang control. According to the
aforementioned analysis in the section, we can choose the piecewise
constant controls $\{f(t):0\leq{t}\leq{t_{f}}\}$ as follows:
\begin{equation}
\label{3-17-1} f(t)=\bigg\{\begin{array}{cc}
M_{ub}, t\in[0,t_{f})& if 0\pm{2k\pi}<(\phi^{H}_{s}-\phi^{H}_{0})<\pi\pm{2k\pi}\\
-M_{ub}, t\in[0,t_{f})& if \pi\pm{2k\pi}\leq(\phi^{H}_{s}-\phi^{H}_{0})<2\pi\pm{2k\pi}
\end{array}
\end{equation}
where $t_{f}=\frac{\phi^{H}_{s0}}{2M_{ub}}$.

Subsequently, our task is to choose $M_{ub}$  to minimize
 the performance (\ref{J1})
where $E(u(t))=|f(t)|^{2}$ and $\lambda>0$.

 After some careful calculations,
 we have
\begin{equation}
\label{2-19-1}\begin{array}{c}
J=\lambda\frac{\phi^{H}_{s0}}{2M_{ub}}+\frac{M_{ub}\phi^{H}_{s0}}{2}\geq\sqrt{\lambda}\phi^{H}_{s0}
\end{array}
\end{equation}
where the equality holds only if $M_{ub}=\sqrt{\lambda}$.

If only bounded Bang-Bang controls with bound $L_{B}$ are
permitted, then the optimal controls are given as:
\begin{equation}
\label{BO1-1} u^{*}_{z}(t)=\bigg\{\begin{array}{cc}
\cos\theta_{u}L_{B}^{*}, t\in[0,t^{*}_{fB})&if 0\pm{2k\pi}\leq(\phi^{H}_{s}-\phi^{H}_{0})<\pi\pm{2k\pi}\\
-\cos\theta_{u}L_{B}^{*}, t\in[0,t^{*}_{fB})&if \pi\pm{2k\pi}\leq(\phi^{H}_{s}-\phi^{H}_{0})<2\pi\pm{2k\pi}
\end{array}
\end{equation}
and
\begin{equation}
\label{BO2-1} u^{*}_{y}(t)=\bigg\{\begin{array}{cc}
\sin\theta_{u}L_{B}^{*}, t\in[0,t^{*}_{fB})&if 0\pm{2k\pi}\leq(\phi^{H}_{s}-\phi^{H}_{0})<\pi\pm{2k\pi}\\
-\sin\theta_{u}L_{B}^{*}, t\in[0,t^{*}_{fB})&if \pi\pm{2k\pi}\leq(\phi^{H}_{s}-\phi^{H}_{0})<2\pi\pm{2k\pi}
\end{array}
\end{equation}
where
$t^{*}_{fB}=\frac{\phi^{H}_{s0}}{2L_{B}^{*}}$
and
$L_{B}^{*}=\min(\sqrt{\lambda},\frac{L_{B}}{\max(\cos\frac{\theta_{u}}{2},\sin\frac{\theta_{u}}{2})})$.
The optimal performance corresponding to bounded Bang-Bang controls
is $J_{B}^{*}=(\frac{\lambda}{2L_{B}^{*}}+\frac{L_{B}^{*}}{2})\phi^{H}_{s0}$. It is interesting to emphasize that optimal
performance  can
be expressed as $J_{B}^{*}=\lambda\cdot{t^{*}_{fB}}+E^{*}_{B}$ with $E_{B}^{*}=\frac{L_{B}^{*}}{2}\phi^{H}_{s0}$, and
$E^{*}_{B}\cdot{t^{*}_{fB}}=\frac{1}{4}(\phi^{H}_{s0})^{2}$
where $\phi^{H}_{s0}$
is independent of $\lambda$.

If unbounded Bang-Bang controls are permitted, then $L_{B}^{*}=\sqrt{\lambda}$,
$t^{*}_{fB}=\frac{\phi^{H}_{s0}}{2\sqrt{\lambda}}$, $E_{B}^{*}=\frac{\sqrt{\lambda}}{2}\phi^{H}_{s0}$, and
$J_{B}^{*}=\sqrt{\lambda}\phi^{H}_{s0}$. Therefore, $E^{*}_{B}\cdot{t^{*}_{fB}}=\frac{1}{4}(\phi^{H}_{s0})^{2}$.

\subsection{$1$-rotation triangle-function controls}
In this subsection, we will explore how to construct one-rotation triangle-function controls
$\{f(t):0\leq{t}\leq{t_{f}}\}$ to achieve the target state from the initial state. We can select
\begin{equation}
\label{tri2-17-1} f(t)=\bigg\{\begin{array}{cc}
u_{T}(t;0,t_{f},M_{ut})&if 0\pm{2k\pi}\leq(\phi^{H}_{s}-\phi^{H}_{0})<\pi\pm{2k\pi}\\
-u_{T}(t;0,t_{f},M_{ut})&if \pi\pm{2k\pi}\leq(\phi^{H}_{s}-\phi^{H}_{0})<2\pi\pm{2k\pi}
\end{array}
\end{equation}
where $t_{f}=\frac{\phi^{H}_{s0}}{M_{ut}}$.
In other words,
$\{u_{z}(t),u_{y}(t)\}$ can be constructed as follows:
\begin{equation}
\label{tri2-18-1} u_{z}(t)=\bigg\{\begin{array}{cc}
u_{T}(t;0,\frac{\phi^{H}_{s0}}{M_{ut}},M_{ut})\cos\theta_{u}&if 0\pm{2k\pi}\leq(\phi^{H}_{s}-\phi^{H}_{0})<\pi\pm{2k\pi}\\
-u_{T}(t;0,\frac{\phi^{H}_{s0}}{M_{ut}},M_{ut})\cos\theta_{u}&if \pi\pm{2k\pi}\leq(\phi^{H}_{s}-\phi^{H}_{0})<2\pi\pm{2k\pi}
\end{array}
\end{equation}
and
\begin{equation}
\label{tri2-18-2} u_{y}(t)=\bigg\{\begin{array}{cc}
u_{T}(t;0,\frac{\phi^{H}_{s0}}{M_{ut}},M_{ut})\cos\theta_{u}&if 0\pm{2k\pi}\leq(\phi^{H}_{s}-\phi^{H}_{0})<\pi\pm{2k\pi}\\
-u_{T}(t;0,\frac{\phi^{H}_{s0}}{M_{ut}},M_{ut})\cos\theta_{u}&if \pi\pm{2k\pi}\leq(\phi^{H}_{s}-\phi^{H}_{0})<2\pi\pm{2k\pi}
\end{array}
\end{equation}

Next, our task is to optimize magnitude $M_{ut}$
in terms of
 the performance (\ref{J1}). It is easy to demonstrate that
\begin{equation}
\label{Tri2-19-1}\begin{array}{c}
J=\lambda\frac{\phi^{H}_{s0}}{M_{ut}}
+\frac{M_{ut}\phi^{H}_{s0}}{3}
\geq\frac{2\sqrt{\lambda}}{\sqrt{3}}\phi^{H}_{s0}
\end{array}
\end{equation}
where the equality holds only if
$M_{ut}=\sqrt{3\lambda}$.

If only bounded triangle-function controls with bound $L_{B}$
are permitted, then the optimal controls are given as:
\begin{equation}
\label{btriuo1-1}u^{*}_{z}(t)=\bigg\{\begin{array}{cc}
u_{T}(t;0,t^{*}_{fT},L_{T}^{*})\cos\theta_{u}&if 0\pm{2k\pi}\leq(\phi^{H}_{s}-\phi^{H}_{0})<\pi\pm{2k\pi}\\
-u_{T}(t;0,t^{*}_{fT},L_{T}^{*})\cos\theta_{u}&if \pi\pm{2k\pi}\leq(\phi^{H}_{s}-\phi^{H}_{0})<2\pi\pm{2k\pi}
\end{array}
\end{equation}
and
\begin{equation}
\label{btriuo2-1} u^{*}_{y}(t)=\bigg\{\begin{array}{cc}
u_{T}(t;0,t^{*}_{fT},L_{T}^{*})\sin\theta_{u}&if\ 0\pm{2k\pi}\leq(\phi^{H}_{s}-\phi^{H}_{0})<\pi\pm{2k\pi}\\
-u_{T}(t;0,t^{*}_{fT},L_{T}^{*})\sin\theta_{u}&if\  \pi\pm{2k\pi}\leq(\phi^{H}_{s}-\phi^{H}_{0})<2\pi\pm{2k\pi}
\end{array}
\end{equation}
where $t^{*}_{fT}=\frac{\phi^{H}_{s0}}{L_{T}^{*}}$ and $L_{T}^{*}=\min(\sqrt{3\lambda},\frac{L_{B}}{\max(\cos\frac{\theta_{u}}{2},\sin\frac{\theta_{u}}{2})})$.
The optimal performance corresponding to bounded  control is
$J_{T}^{*}=(\frac{\lambda}{L_{T}^{*}}+\frac{L_{T}^{*}}{3})\phi^{H}_{s0}$. It is interesting to emphasize that optimal
performance corresponding to  bounded  triangle-function controls can
be expressed as $J_{T}^{*}=\lambda\cdot{t^{*}_{fT}}+E^{*}_{T}$ with $E_{T}^{*}=\frac{1}{3}L_{T}^{*}\phi^{H}_{s0}$, and
$E^{*}_{T}\cdot{t^{*}_{fT}}=\frac{1}{3}(\phi^{H}_{s0})^{2}$
where $\phi^{H}_{s0}$
is independent of $\lambda$.

 If unbounded triangle-function controls are permitted, then $t^{*}_{fT}=\frac{\phi^{H}_{s0}}{\sqrt{3\lambda}}$,
 $L_{T}^{*}={\sqrt{3\lambda}}$, $E_{T}^{*}=\frac{\sqrt{3\lambda}}{3}\phi^{H}_{s0}$ and $J_{T}^{*}=\frac{2\sqrt{\lambda}}{\sqrt{3}}\phi^{H}_{s0}$.
 Therefore $E^{*}_{T}\cdot{t^{*}_{fT}}=\frac{1}{3}(\phi^{H}_{s0})^{2}$.

\subsection{$1$-rotation quadratic-function controls}
In this subsection, we will explore how to construct quadratic controls
$\{f(t):0\leq{t}\leq{t_{f}}\}$ to achieve the target state from the initial state. We can choose
\begin{equation}
\label{square2-17-1} f(t)=\bigg\{\begin{array}{cc}
u_{Q}(t;0,t_{f},M_{uq})&if 0\pm{2k\pi}\leq(\phi^{H}_{s}-\phi^{H}_{0})<\pi\pm{2k\pi}\\
-u_{Q}(t;0,t_{f},M_{uq})&if \pi\pm{2k\pi}\leq(\phi^{H}_{s}-\phi^{H}_{0})<2\pi\pm{2k\pi}
\end{array}
\end{equation}
where $t_{f}=\frac{3\phi^{H}_{s0}}{4M_{uq}}$. In other words, the quadratic controls
$\{u_{z}(t),u_{y}(t)\}$ are given as follows:
\begin{equation}
\label{sq2-17-1} u_{z}(t)=\bigg\{\begin{array}{cc}
u_{Q}(t;0,\frac{3\phi^{H}_{s0}}{4M_{uq}},M_{uq})\cos\theta_{u}&if 0\pm{2k\pi}\leq(\phi^{H}_{s}-\phi^{H}_{0})<\pi\pm{2k\pi}\\
-u_{Q}(t;0,\frac{3\phi^{H}_{s0}}{4M_{uq}},M_{uq})\cos\theta_{u}&if \pi\pm{2k\pi}\leq(\phi^{H}_{s}-\phi^{H}_{0})<2\pi\pm{2k\pi}
\end{array}
\end{equation}
and
\begin{equation}
\label{sq2-18-1} u_{y}(t)=\bigg\{\begin{array}{cc}
u_{Q}(t;0,\frac{3\phi^{H}_{s0}}{4M_{uq}},M_{uq})\sin\theta_{u}&if 0\pm{2k\pi}\leq(\phi^{H}_{s}-\phi^{H}_{0})<\pi\pm{2k\pi}\\
-u_{Q}(t;0,\frac{3\phi^{H}_{s0}}{4M_{uq}},M_{uq})\sin\theta_{u}&if \pi\pm{2k\pi}\leq(\phi^{H}_{s}-\phi^{H}_{0})<2\pi\pm{2k\pi}
\end{array}
\end{equation}

Next, our task is to choose magnitude $M_{uq}$
to minimize
 the performance (\ref{J1}).
After some calculations,  we further obtain
\begin{equation}
\label{Sqi2-19}\begin{array}{c}
J=\lambda\frac{3\phi^{H}_{s0}}{4M_{uq}}
+\frac{2M_{uq}\phi^{H}_{s0}}{5}\geq\frac{{\sqrt{30\lambda}}}{5}\phi^{H}_{s0}
\end{array}
\end{equation}
where the equality holds only if
$M_{uq}=\frac{\sqrt{30\lambda}}{4}$.

If only bounded quadratic controls with bound $L_{B}$ are
permitted, then the optimal controls are given as:
\begin{equation}
\label{bsqruo1-1}u^{*}_{z}(t)=\bigg\{\begin{array}{cc}
u_{Q}(t;0,t^{*}_{fQ},L_{Q}^{*})\cos\theta_{u}&if 0\pm{2k\pi}\leq(\phi^{H}_{s}-\phi^{H}_{0})<\pi\pm{2k\pi}\\
-u_{Q}(t;0,t^{*}_{fQ},L_{Q}^{*})\cos\theta_{u}&if \pi\pm{2k\pi}\leq(\phi^{H}_{s}-\phi^{H}_{0})<2\pi\pm{2k\pi}
\end{array}
\end{equation}
and
\begin{equation}
\label{bsqruo2-1} u^{*}_{y}(t)=\bigg\{\begin{array}{cc}
u_{Q}(t;0,t^{*}_{fQ},L_{Q}^{*})\sin\theta_{u}&if 0\pm{2k\pi}\leq(\phi^{H}_{s}-\phi^{H}_{0})<\pi\pm{2k\pi}\\
-u_{Q}(t;0,t^{*}_{fQ},L_{Q}^{*})\sin\theta_{u}&if \pi\pm{2k\pi}\leq(\phi^{H}_{s}-\phi^{H}_{0})<2\pi\pm{2k\pi}
\end{array}
\end{equation}
where $t^{*}_{fQ}=\frac{3\phi^{H}_{s0}}{4L_{Q}^{*}}$ and
$L_{Q}^{*}=\min(\frac{\sqrt{30\lambda}}{4},\frac{L_{B}}{\max(\cos\frac{\theta_{u}}{2},\sin\frac{\theta_{u}}{2})})$.
The optimal performance corresponding to bounded  control is
$J_{Q}^{*}=(\frac{3\lambda}{4L_{Q}^{*}}+\frac{2L_{Q}^{*}}{5})(\phi^{H}_{s0})$.
It is interesting to emphasize that optimal performance
corresponding to unbounded  quadratic controls can be expressed as
$J_{Q}^{*}=\lambda\cdot{t^{*}_{fQ}}+E^{*}_{Q}$ with
$E_{Q}^{*}=\frac{2}{5}L_{Q}^{*}\phi^{H}_{s0}$, and
$E^{*}_{Q}\cdot{t^{*}_{fQ}}=\frac{3}{10}(\phi^{H}_{s0})^{2}$ where
$\phi^{H}_{s0}$ is independent of $\lambda$.

If unbounded quadratic controls are permitted, then  $t^{*}_{fQ}=\frac{3\phi^{H}_{s0}}{\sqrt{30\lambda}}$,
 $L_{Q}^{*}=\frac{\sqrt{30\lambda}}{4}$,
 $E_{Q}^{*}=\frac{{\sqrt{30\lambda}}}{10}\phi^{H}_{s0}$,
$J_{Q}^{*}=\frac{\sqrt{30\lambda}}{5}\phi^{H}_{s0}$,  and
$E^{*}_{Q}\cdot{t^{*}_{fQ}}=\frac{3}{10}(\phi^{H}_{s0})^{2}$.

\subsection{Further discussions}
1. When unbounded controls are permitted, we have
$J^{*}_{B}<J^{*}_{Q}<J^{*}_{T}$, $E^{*}_{B}<E^{*}_{Q}<E^{*}_{T}$ and
 $t^{*}_{fB}<t^{*}_{fQ}<t^{*}_{fT}$, therefore we have $t^{*}_{fB}{\cdot}E^{*}_{B}<t^{*}_{fQ}{\cdot}E^{*}_{Q}<t^{*}_{fT}{\cdot}E^{*}_{T}$.


2. Even when only bounded controls are permitted, the aforementioned
inequalities are valid for all $\lambda$ and $L_{B}$ except that the
inequality $E^{*}_{B}<E^{*}_{Q}<E^{*}_{T}$
 is invalid  for some
$\lambda$ and $L_{B}$.

3. When one fixed Hamiltonian and another tunable control Hamiltonian
are available,  only $1-$rotation  Bang-Bang control can be designed
to transfer the qubit from the initial state to the target state.
For example, if
$H(t)=(\sigma_{z}+u_{y}(t)\sigma_{y})|\psi(t)\rangle$ and
$\sin\theta_{0}\sin\phi_{0}\neq\sin\theta_{s}\sin\phi_{s}$, one may
be able to construct $1-$rotation  Bang-Bang control to achieve the
target state. When unbounded Bang-Bang control are available, one
should choose   $u_{y}(t)=\tan\theta_{u}$ where
$\tan\theta_{u}=\frac{\cos\theta_{s}-\cos\theta_{0}}{\sin\theta_{0}\sin\phi_{0}-\sin\theta_{s}\sin\phi_{s}}$.
When only bounded Bang-Bang controls with the bound $L_{B}$ are
available,  $1-$rotation  bounded Bang-Bang control can be
constructed only if $L_{B}\geq{|\tan\theta_{u}|}$. This result is in
interesting contrast with the recent research\cite{d53}.

\section{Discussions and conclusions}
At first, we would like to point out that  the three-rotation and
one-rotation control design methods can be generalized to manipulate
encoded qubit on controllable subspace of both closed and open
quantum systems.

For example, let us consider a controlled $2$-qubit system which is
governed by the equation
\begin{equation}
\label{3-13}
\frac{d}{dt}|\psi(t)\rangle=-{\frac{i}{\hbar}}H(u(t))|\psi(t)\rangle
\end{equation}
where
$H(u(t))=u_{z_{1}I_{2}}(t)\sigma^{(1)}_{z}\otimes{I^{(2)}_{2}}+u_{I_{1}z_{2}}(t)I^{(1)}_{2}\otimes{\sigma^{(2)}_{z}}
+u_{y_{1}x_{2}}(t)\sigma^{(1)}_{y}\otimes{\sigma^{(2)}_{x}}+u_{x_{1}y_{2}}(t)\sigma^{(1)}_{x}\otimes{\sigma^{(2)}_{y}}$.
Under the above condition, an encoded qubit basis can be given as
$\{|0_{L}\rangle=|0_{1}1_{2}\rangle,|1_{L}\rangle=|1_{1}0_{2}\rangle\}$.
Denote the encoded subspace, which can be expanded by the encoded
state basis $\{|0_{L}\rangle,|1_{L}\rangle\}$,   as $E_{L}$.  It is
interesting to underline that for any pure state
$|\psi_{E}\rangle{\in}E_{L}$, one can obtain its geometric
parametrization in terms of $\{|0_{L}\rangle=|0_{1}1_{2}\rangle$ and
$|1_{L}\rangle=|1_{1}0_{2}\rangle\}$. Denote
$\sigma^{L}_{z}=|0_{L}\rangle\langle0_{L}|-|1_{L}\rangle\langle1_{L}|=\frac{1}{2}(\sigma^{(1)}_{z}\otimes{I^{(2)}_{2}}-{I^{(1)}_{2}}\otimes\sigma^{(2)}_{z})$
and
$\sigma^{L}_{y}=i|1_{L}\rangle\langle0_{L}|-i|0_{L}\rangle\langle1_{L}|=\frac{1}{2}(\sigma^{(1)}_{y}\otimes{\sigma^{(2)}_{x}}-\sigma^{(1)}_{x}\otimes\sigma^{(2)}_{y})$
By setting
$u_{z_{1}I_{2}}(t)=-u_{I_{1}z_{2}}(t)=\frac{1}{2}u^{L}_{z}(t)$ and
$u_{y_{1}x_{2}}(t)=-u_{x_{1}y_{2}}(t)=\frac{1}{2}u^{L}_{y}(t)$, one
can express  the equation (\ref{3-13}) as
\begin{equation}
\label{3-14}
\frac{d}{dt}|\psi(t)\rangle=-{{\frac{i}{\hbar}}}(u^{L}_{z}(t)\sigma^{L}_{z}+u^{L}_{y}(t)\sigma^{L}_{y})|\psi(t)\rangle
\end{equation}

For an open quantum system, its dynamics equation is in general
rather difficult to gain. However, in many practical situation,
quantum dynamical semi-group master
equation\cite{Lindblad-30,Alicki-31} is an appropriate way to
describe the evolution of the quantum open system as follows
\begin{equation}
\label{E4}
\frac{\partial\rho}{\partial{t}}=-\frac{i}{\hbar}[{H(u(t))},\rho]+L(\rho)
\end{equation}
where Lindbladian is:
\begin{equation}
\label{E5}
L(\rho)=\frac{1}{2}\sum_{i,j}^{N}\alpha_{ij}([F_{i},\rho{F^{+}_{j}}]+[F_{i}\rho,F^{+}_{j}])
\end{equation}
and  ${H(u(t))}$ is the system Hamiltonian, the operators $F_{i}$
constitute a basis for the $N$-dimensional space of all bounded
operators acting on $H$, and $\alpha_{ij}$ are the elements of a
positive semi-definite Hermitian matrix.

If
$\hat{H}(u(t))=u_{z_{1}I_{2}}(t)\sigma^{(1)}_{z}\otimes{I^{(2)}_{2}}+u_{I_{1}z_{2}}(t)I^{(1)}_{2}\otimes{\sigma^{(2)}_{z}}
+u_{y_{1}x_{2}}(t)\sigma^{(1)}_{y}\otimes{\sigma^{(2)}_{x}}+u_{x_{1}y_{2}}(t)\sigma^{(1)}_{x}\otimes{\sigma^{(2)}_{y}}$
and  $L(|\psi_{E}\rangle\langle\psi_{E}|)=0$ for any pure state
$|\psi_{E}\rangle{\in}E_{L}$, then,  for
$\rho=|\psi_{E}\rangle\langle\psi_{E}|$ with
$|\psi_{E}\rangle{\in}E_{L}$,   Eq.(\ref{E4}) is further reduced to
Eq. (\ref{3-14}) because $L(|\psi_{E}\rangle\langle\psi_{E}|)=0$.

So far, it has been demonstrated in this research that one can
utilize various local wave-function controls including Bang-Bang
controls, triangle-function controls and quadratic-function controls
to manipulate qubits and encoded qubits on controllable subspaces
for both open quantum dynamical systems and uncontrollable closed
quantum dynamical systems when two tunable Hamiltonian controls are
available. Furthermore, we discuss how to design control magnitude
in terms of a kind of time-energy performance. It is demonstrated
that optimal Bang-Bang controls have the best performance and
optimal triangle-function controls have the worst performance among
three kinds of control schemes. It is   the pulse area inequality
for three controls given in Eq. (\ref{Area-i}) who makes the
performance difference. It should be emphasized that one can
introduce a ratio parameter $\lambda$ to trade-off between time and
energy resource cost, but the product of time and energy cost is an
invariance under different $\lambda$ for each kind of controls due
to the characteristic of geometric control.

It is well known that low-capacitance Josephson tunneling junctions
offer a promising way to realize qubits for quantum information
processing\cite{J} and two tunable Hamiltonian controls are
available in this application. Therefore this research implies that
one can constructively adjust gate voltages or magnetic fields to
manipulate qubits based on either charge or phase (flux) degrees of
freedom .



%


\end{document}